\documentclass[twocolumn,pre,amsmath,amssymb,superscriptaddress,natbib]{revtex4-1}
\usepackage{blindtext}
\usepackage{graphicx}
\usepackage{epstopdf}
\usepackage{epsfig}
\usepackage{dcolumn}
\usepackage{bm}
\usepackage{color}
\usepackage{subfigure}
\usepackage{wrapfig}
\usepackage{rotate,color,url}
\usepackage{longtable}
\usepackage[pagebackref=false,colorlinks,linkcolor=blue,citecolor=blue,urlcolor=magenta]{hyperref}

\begin{document}
\newcommand {\be}{\begin{equation}}
\newcommand {\ee}{\end{equation}}
\newcommand {\bea}{\begin{array}}
\newcommand {\cl}{\centerline}
\newcommand {\eea}{\end{array}}
\newcommand {\pa}{\partial}
\newcommand {\al}{\alpha}
\newcommand {\de}{\delta}
\newcommand {\ta}{\tau}
\newcommand {\ga}{\gamma}
\newcommand {\ep}{\epsilon}
\newcommand {\si}{\sigma}
\newcommand{\up}{\uparrow}
\newcommand{\down}{\downarrow}
\title{Spike-Phase Coupling as an Order Parameter in a Leaky Integrate-and-Fire Model}
\author{Nahid Safari}
\affiliation{Department of Physics, Isfahan University of Technology, Isfahan, Iran}

\author{Farhad Shahbazi}
\affiliation{Department of Physics, Isfahan University of Technology, Isfahan, Iran}

\author{Mohammad Dehghani-Habibabadi}
\affiliation{Department of Neurophysics, Institute for Theoretical Physics, University of Bremen, Bremen, Germany}

\author{Moein Esghaei}
\affiliation{Cognitive Neuroscience Laboratory, German Primate Center- Leibniz Institute for Primate Research, G\"{o}ttingen, Germany}

\author{\thanks{corresponding author}Marzieh Zare}
\affiliation{D\'{e}partement de psychologie, Universit\'{e} de Montr\'{e}al, Montr\'{e}al, Canada}
\email{marzieh.zare@umontreal.ca}

\date{\today}
\begin{abstract}
While criticality is widely observed in neural networks, its underlying neural mechanism is not known well. We consider a network of $N$ excitatory leaky integrated and fire (LIF) neurons that reside on a regular lattice with periodic boundary conditions. The cooperation between neurons, $K$, plays the role of the control parameter that is expected to generate criticality when the critical cooperation strength, $K_c$, is adopted. We show that the coupling between spike timing and the phase of temporal fluctuations of a cooperative activity of the network, i.e. population-averaged voltage (PAV), resorts to identifying an order parameter. By increasing $K$, we find a continuous transition from irregular spiking to a phase-locked state at the critical point, $K_c$. Moreover, we deploy the finite-size scaling analysis to obtain the critical exponents of this transition. We also show that the neuronal avalanches created at this critical point, display a remarkable scaling behavior with the exponents in a fair agreement with the experimental values.
\end{abstract}
\keywords{Spike phase-coupling, Population-averaged voltage, Criticality, Order parameter, Neural avalanches}
\maketitle
\section{Introduction\label{Intro}}

Understanding the mechanisms that underlie, or are generated via criticality in neuronal networks are of particular interest~\cite{schuster2014criticality, munoz2018colloquium}. The question is ``under what conditions do local cortical networks converge on criticality or deviate from criticality?" Royer and Pare~\cite {royer2003conservation} proposed that conservation of synaptic strengths is a possible mechanism responsible for maintaining a network in the vicinity of a critical point. Similarly, in~\cite{hsu2006d}, authors show that due to some combination of homeostatic regulation of excitability and Hebbian learning, the total sum of synaptic strengths remains near a constant value, hence the network operates near criticality. Competitive activity-dependent attachment and pruning~\cite{bornholdt2000s,de2006self}, dynamic synaptic plasticity~\cite{levina2007dynamical}, synaptic scaling~\cite{turrigiano2000hebb}, and neural coupling adjustment to generate criticality~\cite{Zare2013,Zare2012,Zare2015,Dehghani2017} are possible mechanisms for leading up the network into critical regime. 

Ongoing activity in the cortex results from activities of many single neurons that transiently add up to larger events, traceable in various recording techniques including extracellular microelectrode arrays (MEAs) with local field potential (LFP), the electroencephalogram (EEG), the magnetoencephalogram (MEG), and the BOLD signal from functional magnetic resonance imaging (fMRI)~\cite{plenz2011multi}. An attractive yet controversial dynamical landscape explores emergent collective behavior in the brain, proposing the concept of ``neuronal avalanches". Different from avalanches in critical sandpile models which generate stochastic patterns, neural avalanches occur in spatiotemporal patterns, display stable scale-free patterns in the spatial extent of neuronal activity, i.e. patterns with temporal {\it precision} of $\sim 4 ms$ are reproducible over periods of as long as 10 hours~\cite{Beggs2003,beggs2004neuronal}. These properties suggest that avalanches could be used by networks as a substrate for storing information ~\cite{beggs2007criticality,Socolar2003}. Moreover, neural avalanche patterns are considered to be a signature of criticality, and a lot of empirical evidence supports the hypothesis that the cortex operates near criticality, thereby enabling efficient information processing~\cite{Beggs2003,Beggs2012,Friedman2012,Yang2012,Socolar2003,Shew2009}, dynamic range~\cite{Shew2009,Shew2011,kinouchi2006optimal}, communication and information transition~\cite{Beggs2003}, optimal communication~\cite{Beggs2003,Beggs2012,Bertschinger2004,Rama2007,Shew2011,Tanaka2009,Chialvo2010}, transition capability ~\cite{Shew2011}, computational power~\cite{Bertschinger2004}, phase synchrony~\cite{Yang2012} and maximizing brain functional diversity ~\cite{wang2019hierarchical}.

For about two decades, it was believed that the observed criticality in neuronal avalanches is a branching process and its universality is governed by the mean-filed directed percolation (MF-DP). In this scenario, the brain operates at the vicinity of an absorbing-active critical point~\cite{schuster2014criticality}.
However, recent works support the idea that the transition governing brain dynamics is indeed between active and oscillating phases, rather than being between absorbing and active phases~\cite{Yang2012,poil2012critical,di2018landau,dalla2019modeling,fontenele2019criticality}.

Despite remarkable progress in understanding the criticality in neuronal systems, the definition of a suitable order parameter for a description of such critical behavior is still missing.  Motivated by the above recent results, in this work we define an order parameter for quantifying the desynchronized-synchronized transition in a leaky integrate-and-fire (LIF) network.  To this end, we study the coupling between sequences of spikes~\cite{abeles1993spatiotemporal,hahnloser2002ultra}, and the phase of temporal fluctuations of a macroscopic observable of the networks, that is the population-average voltage (PAV), in the model we have previously introduced~\cite{Zare2013,Zare2012,Zare2015,Dehghani2017}. We explore whether such a spike-phase coupling (SPC) can detect the different magnitude of network-level synchrony, thus playing the role of an order parameter in the model. SPC is a proxy to capture synchrony of single-neuron activity with accumulated electric current from all active neurons within the network. Previously, using the temporal complexity approach relying on events generated in time, we introduced a robust indicator of criticality~\cite{Zare2013,Zare2012} in the form of a function called Mittag-Leffler (ML) function. Despite its robustness in the detection of criticality compared to neural avalanches~\cite{Dehghani2017}, the lack of a dynamical theory for the origin of the ML function led to the absence of a theory and a robust {\it order parameter} for the form of criticality generated by the neural model. Here, however, using SPC quantified by Phase-Locking Value (PLV), we illustrate that by increasing the coupling (connection strength, $K$) between neurons, a second-order transition from desynchronized firing with low SPC to a synchronized firing state with high SPC is realized at the critical value, $K_c$. This suggests that SPC plays the role of an order parameter in the model. Moreover, using the finite-size scaling analysis, we evaluate the critical exponents incorporating in this transition.
We also evaluate the scaling exponents of the neural avalanches of the model at the critical point and show that they are in good agreement with the recent experimental results~\cite{fontenele2019criticality}.

The paper is organized as follows. The neuronal model and simulation details are introduced in section~\ref{Model}. Section~\ref{SPC} discusses the spike-PAV phase coupling and its role as the order parameter of a critical transition in the model. The scaling of neuronal avalanches are studied in section~\ref{Scaling}, and section~\ref{conclusion} is devoted to the concluding remarks.


\section{Model Description \label{Model}}

$N$ excitatory neurons reside on a two-dimensional square lattice with the periodic boundary condition of size $L$, where $N = L \times L$. We use the leaky integrate-and fire model (LIFM) ~\cite{Politi} as used in our previous studies~\cite{Zare2013,Zare2012,Zare2015,Dehghani2017}

\begin{equation} \label{LIFM}
\dot{x_i} = - \gamma x_{i}(t) + S + \sigma \xi_i (t),
\end{equation}

\noindent where $x$ is the membrane potential, $1/\gamma$ is the membrane time constant of the neuron and $S$ denotes the constant input current. $\xi(t)$ is a continuous Gaussian white noise with zero mean and unit variance, defined by


\begin{eqnarray} \label{noise}
\langle \xi(t) \rangle &=&0 \nonumber\\
\langle \xi(t) \xi(t') \rangle &=& \delta(t-t'),
\end{eqnarray}

\noindent The parameter $\sigma$ is considered as the noise intensity. Starting from a random value or zero, we consider $x = 1$ as the firing threshold for each neuron. After reaching the threshold, the neuron adds a value $K$ to the input currents of its neighboring neurons and jumps back to a rest state with $0<x<0.5$. The quantity $K>0$ implies that all the neurons linked to a firing neuron, make a step forward by $K$ which is called the cooperation strength. This parameter plays the role of the control parameter, and we later refer to it as the generator of criticality when $K_c$, critical cooperation strength is adopted. Therefore, in our analysis, we investigate the system's behavior while we vary the coupling strength for each realization. It is noteworthy that the model generates spikes simply extracted from numerical analysis.

\noindent For the numerical analyses, we chose $S = 0.001005$, $\gamma = 0.001$, and $\sigma=0.0001$ and the lattices with the linear sizes of $L = {20, 25, 30, 35,40}$. We adopted the integration time step $\Delta t= 0.01$, and the duration of analysis were $10^7$ time steps. Computations were carried out using the Python programming language (Python Software Foundation. Python Language Reference, version 3.7.2. available at \url{https://www.python.org/}).

\section{Spike-phase coupling and criticality in neuronal systems\label{SPC}}

\begin{figure}
\centering
\resizebox{0.9\columnwidth}{!}{%
\includegraphics{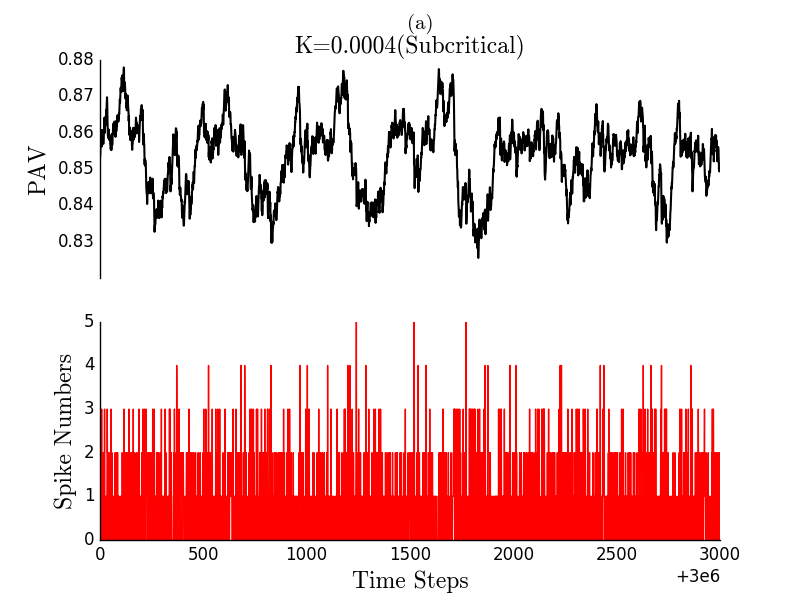}}
\resizebox{0.9\columnwidth}{!}{%
\includegraphics{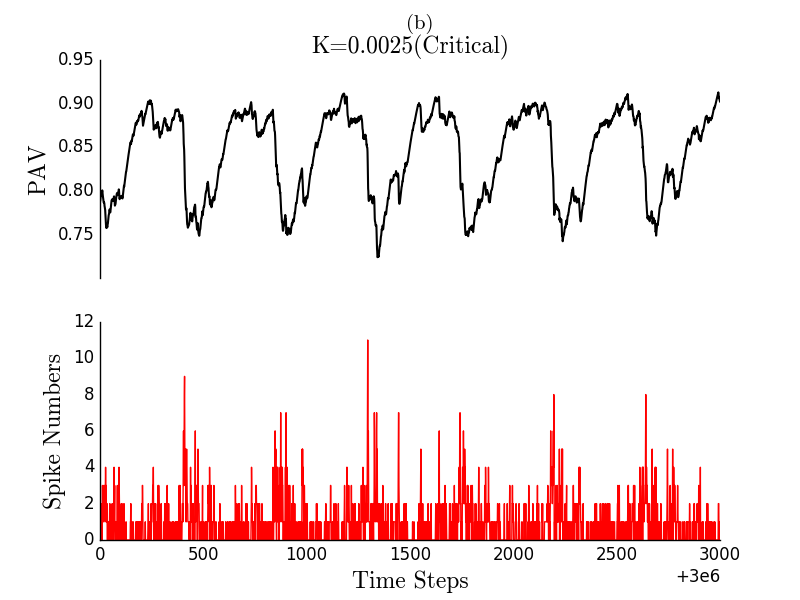}}
\resizebox{0.9\columnwidth}{!}{%
\includegraphics{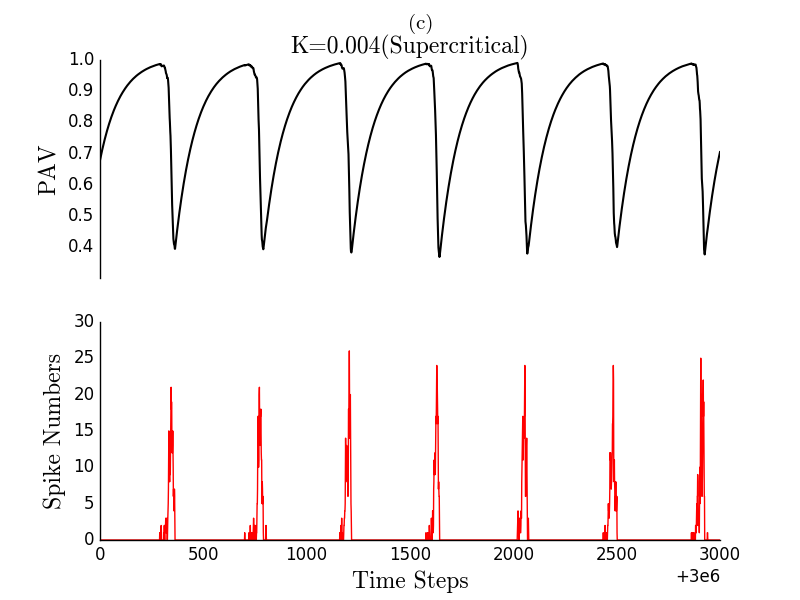}}
\caption{ (Color online) Spike and PAV time series for (a) subcritical regime, $K=0.0004$, (b) critical regime, $K=0.0025$. (c) supercritical regime, $K=0.004$.} 
\label{fig:spc}
\end{figure}

\begin{figure}
\centering
\resizebox{\columnwidth}{!}{%
\includegraphics{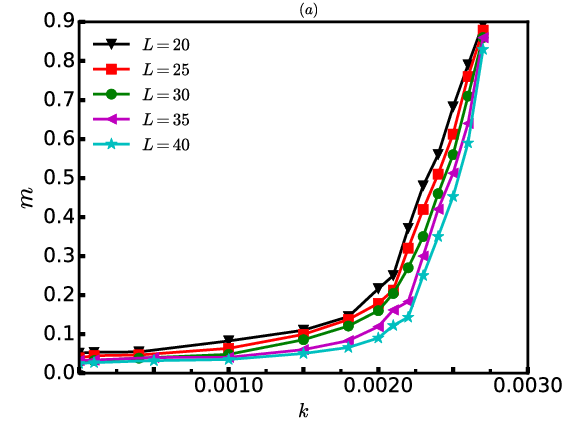}}
\resizebox{\columnwidth}{!}{%
\includegraphics{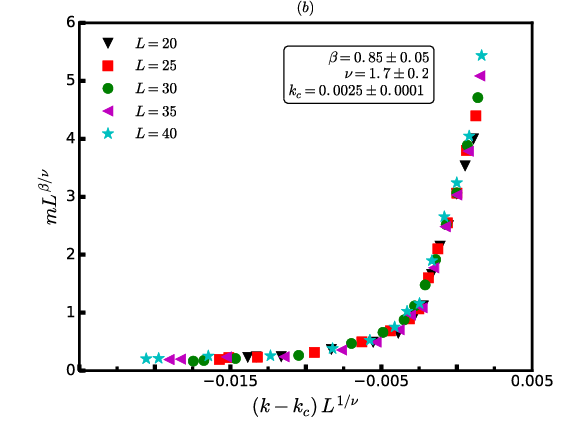}}
\caption{ (Color online) (a) Dependence of absolute value of PLV ($m$) on the parameter $K$ for lattices with the linear size $L=20,25,30,35,40$. (b) Collapse of $m-K$ plots for different system sizes as the result of rescaling.}
\label{fig:plv}
\end{figure}

Here, we quantify SPC as reflective of the coupling between the phase of PAV and neuronal spike timing. PAV is the average of neuronal potentials over all active neurons~\cite{golomb1994clustering,golomb2000number}, hence at each time step we calculate it as ${\rm PAV}(t) = \frac{1}{N}\sum_{i=1}^N x_{i}(t)$, where $x_{i}(t)$ is the membrane voltage of the $i^{th}$ neuron at the time $t$. Fig. \ref{fig:spc} illustrates the PAV and number of spikes (summed across the population of neurons) for three values of parameter $K=0.0004$ (subrcitical), $K=0.0025$ (critical), $K=0.004$ (supercritical). To this end, for each spike-PAV pair, the parameter $M_j$ is defined as

\begin{equation}
\label{M}
M_{j}=\frac {1}{n} \sum_{k=1}^n \exp{(i\phi^{j}_k)},
\end{equation}
\noindent where $n$ is the number of spikes in each spike train (for every single neuron), $\phi^{j}_k$ is the instantaneous phase of PAV corresponding to the $k^{th}$ spike of the $j^{th}$ neuron and is extracted from the Hilbert transform of the PAV time series. To obtain the phase-locking value (PLV) for each trial, vector $M$ was computed for each spike train and then the mean vector was calculated across them using the following ~\cite{PLV}

\begin{equation}
\label{plv}
{\rm PLV}=\frac{1}{N} \sum_{i=1}^N M_i .
\end{equation}

In the case of no spike coupling, PLV goes to $0$ for large spike trains ($n\rightarrow \infty$), and for full synchrony at which all neurons spike at the same phase of PAV, the absolute value of PLV reaches its maximum $1$. 

Therefore, an estimation of the strength of SPC is given by the absolute value of PLV ($m=|{\rm PLV}|$). Indeed, $m$ can be considered as an order parameter indicating the transition from irregular to synchronous spiking pattern.  The variations of the order parameter $m$, versus cooperation parameter $K$, for $L=20, 25, 30, 35, 40$, illustrated in the top panel of Fig.~\ref{fig:plv} suggests a continuous transition from irregular to synchronous condition as $K$ increases. To verify the continuous transition and critical behavior in this system, we use the finite-size scaling theory. Hence, we assume the following scaling of the dependence of the order parameter on the coupling strength (cooperation), $K$

\begin{equation}
m=L^{-\beta/\nu}{\cal M}(|K-K_{c}|L^{1/\nu})
\end{equation}
where $K_c$ is the value of critical cooperation and $\beta$ and $\nu$ are the critical exponents corresponding to order parameter and correlation length, respectively. ${\cal M}(x)$ is the scaling function, for which one expects the following behavior at large system sizes

\begin{equation}
{\cal M}(x)\sim x^{\beta} ~ {\rm as}~ x\rightarrow \infty.
\end{equation}
This property gives rise to the following behavior for the order parameter in infinite system size limit, for $K>K_c $

\begin{equation}
m\sim (K-K_c)^{\beta},
\end{equation}
which is an indicator of a continuous phase transition.

The bottom panel of Fig.~\ref{fig:plv} clearly shows the data collapse of the different lattice sizes by choosing $K_c=0.0025\pm 0.0001$, $\beta = 0.85 \pm 0.05$ and $\nu= 1.7 \pm 0.2$.  These exponents indicate that the universality of this transition is different from the mean-field directed percolation with the exponents $\beta=1$ and $\nu=1/2$, and also directed percolation in $2+1$ dimensions whose exponents are calculated as $\beta=0.580(4)$ and $\nu=0.729(1)$~\cite{wang2013high}.

\section{Scaling of the neuronal avalanches \label{Scaling}}

\begin{figure}
\centering
\resizebox{\columnwidth}{!}{%
\includegraphics{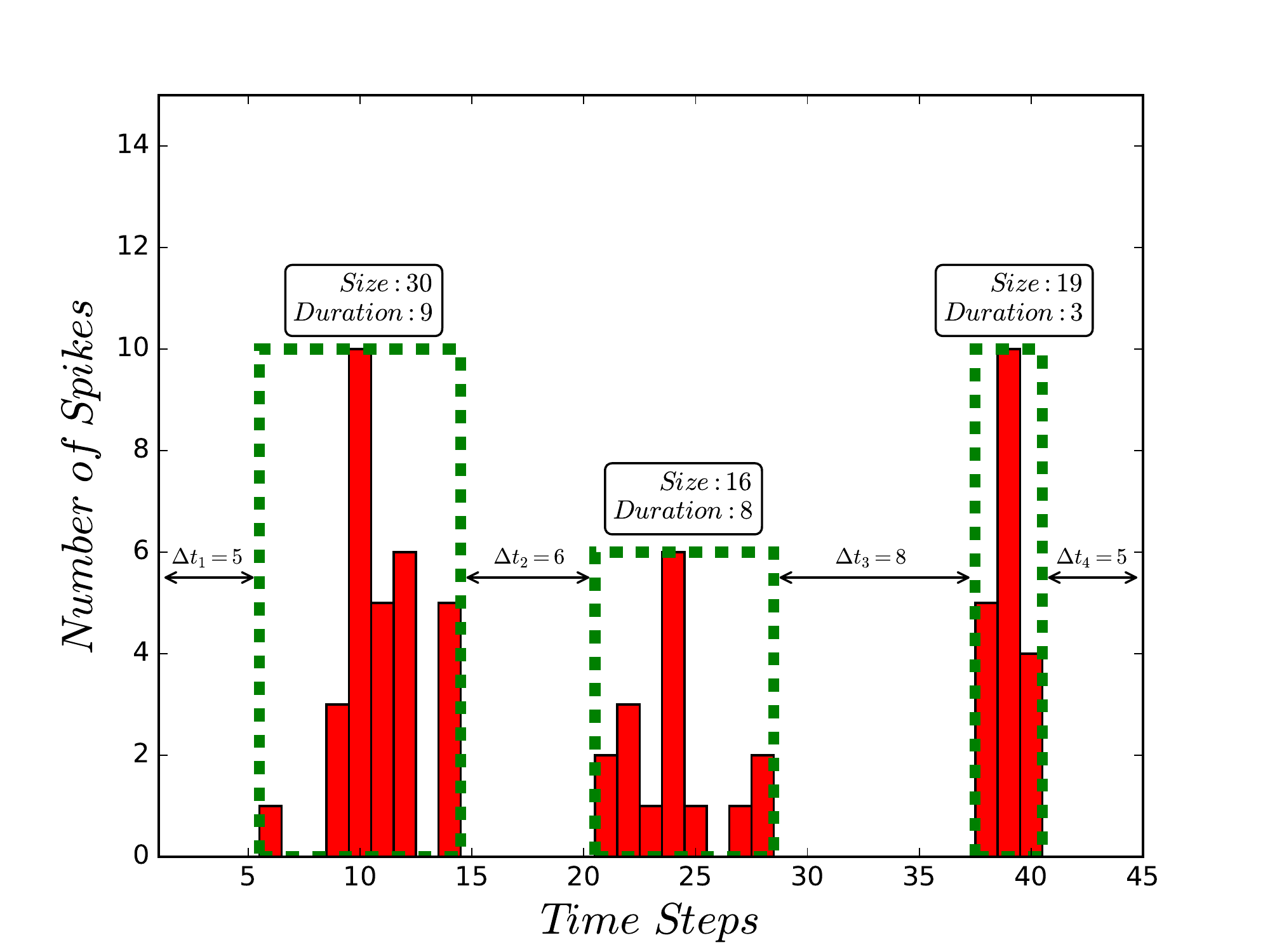}}
\caption{ (Color online) Schematic plot of calculating the neuronal avalanches.}
\label{fig:avalanche_def}
\end{figure}

In our former studies \cite{Zare2013,Zare2012,Dehghani2017}, we have extensively analyzed the emergence of criticality in the LIFM model described above using temporal complexity and avalanche analysis. In this section, we complement our former analyses to prove the existence of criticality using finite-size scaling theory. We are interested to find the special $K_c$ or the narrow region at which the system displays the critical behavior. As documented previously, the emergence of criticality is indicated by the existence of power-law exponent of avalanche size, lifetime distributions limited to a specific range ~\cite{Beggs2003}, and the shape collapse as suggested by Friedman and colleagues, in cultured slices of cortical tissue ~\cite{Friedman2012}. The widely varying profile of neural avalanche distribution in size is described by a single universal scaling exponent, $\tau$ in size, $p(S) \sim S^{-\tau}$, and a single universal exponent in duration, $\alpha$, $p(T) \sim T^{-\alpha}$, where $S$ and $T$ denote the size, and the duration of avalanches, respectively.

\begin{figure}
\centering
\resizebox{\columnwidth}{!}{%
\includegraphics{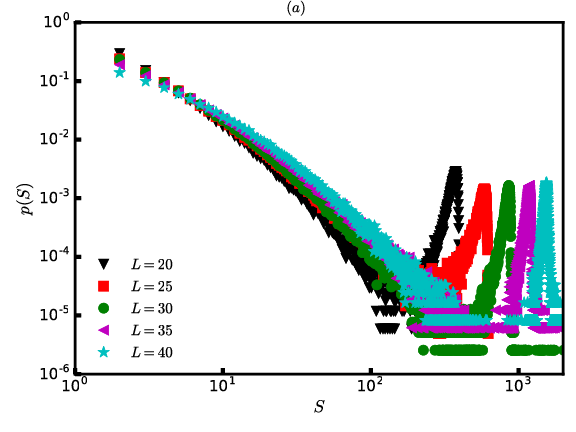}}
\resizebox{\columnwidth}{!}{%
\includegraphics{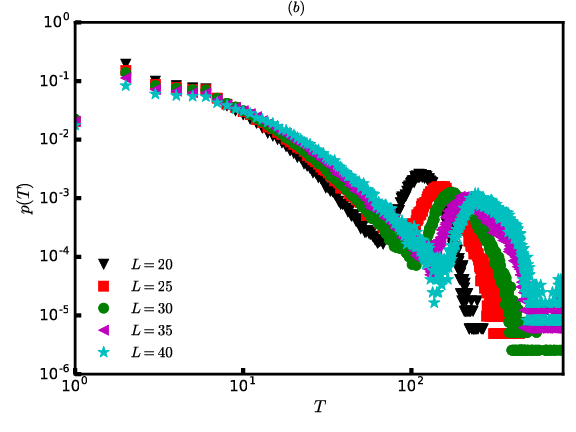}}
\resizebox{\columnwidth}{!}{%
\includegraphics{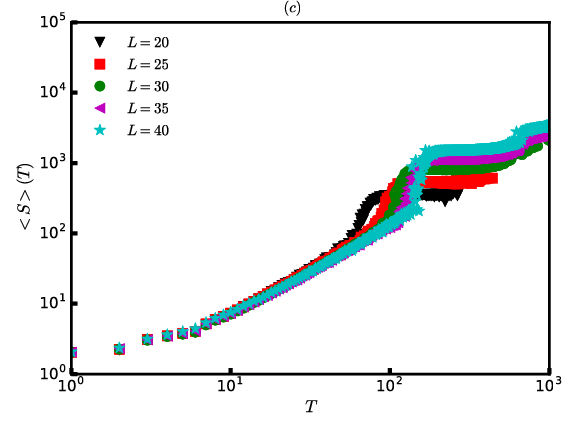}}
\caption{ (Color online) Log-Log plots of the probability density functions of (a) avalanche size $P(S)$ and (b) avalanche duration
$P(T)$ and conditional average of avalanche size versus the duration for $K=0.0025$. The results are obtained on the square lattices with the linear size $L=20,25,30,35,40$.}
\label{fig:avalanches}
\end{figure}
According to the single scaling theory, the avalanche data collapse onto universal scaling functions near a critical point as follows ~\cite{James}:

\begin{eqnarray} \label{fs}
p(S) &=& L^{-\mu\tau}{\cal S}(SL^{-\mu}),\nonumber\\
p(T) &=& L^{-z\mu\alpha}{\cal T}(TL^{-z\mu}),\nonumber\\
\langle S \rangle (T) &=&L^{\mu}{\cal F}(TL^{-z\mu}),
\end{eqnarray}
\noindent where $p(S)$, and $p(T)$ are the probability density functions of the avalanche size and duration, respectively. $\langle S \rangle (T)$ is the average of avalanche size conditioned on a given duration ~\cite{James}. $\tau$ and $\alpha$, are the scaling exponents of the probability density functions of size and duration of the avalanches, respectively. $\mu$ denotes the scaling exponent of the size of avalanches versus the linear size of the lattice and $z$ is the dynamical exponent which determines the scaling relations of the duration and size of the avalanches.

${\cal S}(x)$, ${\cal T}(x)$ and ${\cal F}(x)$ are the universal functions whose behavior at small $x$ limit (large $L$) are as the following
\begin{eqnarray}
\lim_{x\rightarrow 0}{\cal S}(x)&\sim &x^{-\tau}\nonumber\\
\lim_{x\rightarrow 0}{\cal T}(x)&\sim &x^{-\alpha}\nonumber\\
\lim_{x\rightarrow 0}{\cal F}(x)&\sim &x^{1/z},
\end{eqnarray}
and at large $x$ limit (small $L$) each function tends to a constant value. In this way at the infinite lattice size, one finds

\begin{eqnarray}
p(S)&\sim &S^{-\tau}\nonumber\\
p(T)&\sim &T^{-\alpha}\nonumber\\
\langle S \rangle (T)&\sim &T^{1/z},
\end{eqnarray}

\begin{figure}
\centering
\resizebox{0.95\columnwidth}{!}{%
\includegraphics{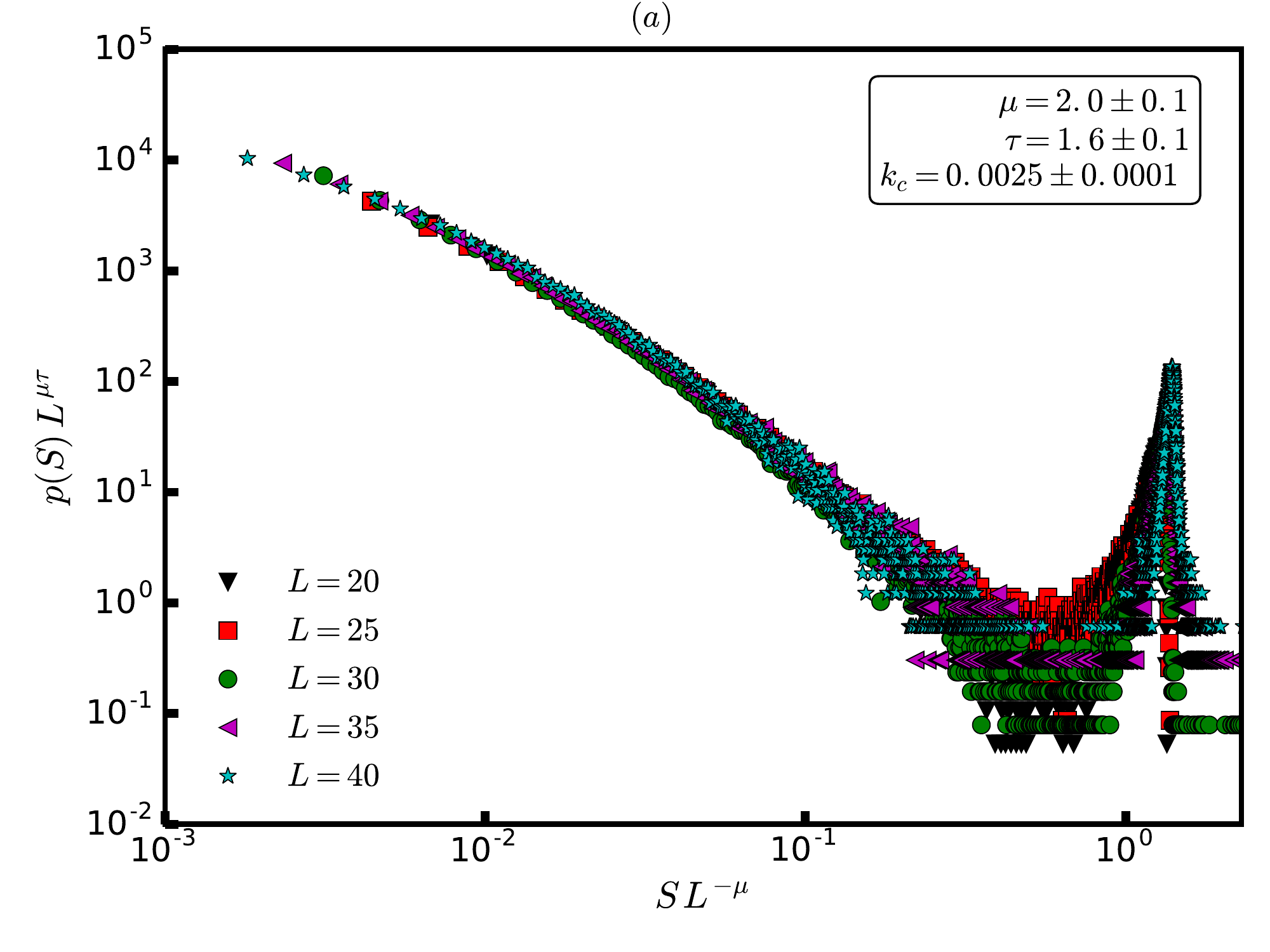}}
\resizebox{0.95\columnwidth}{!}{%
\includegraphics{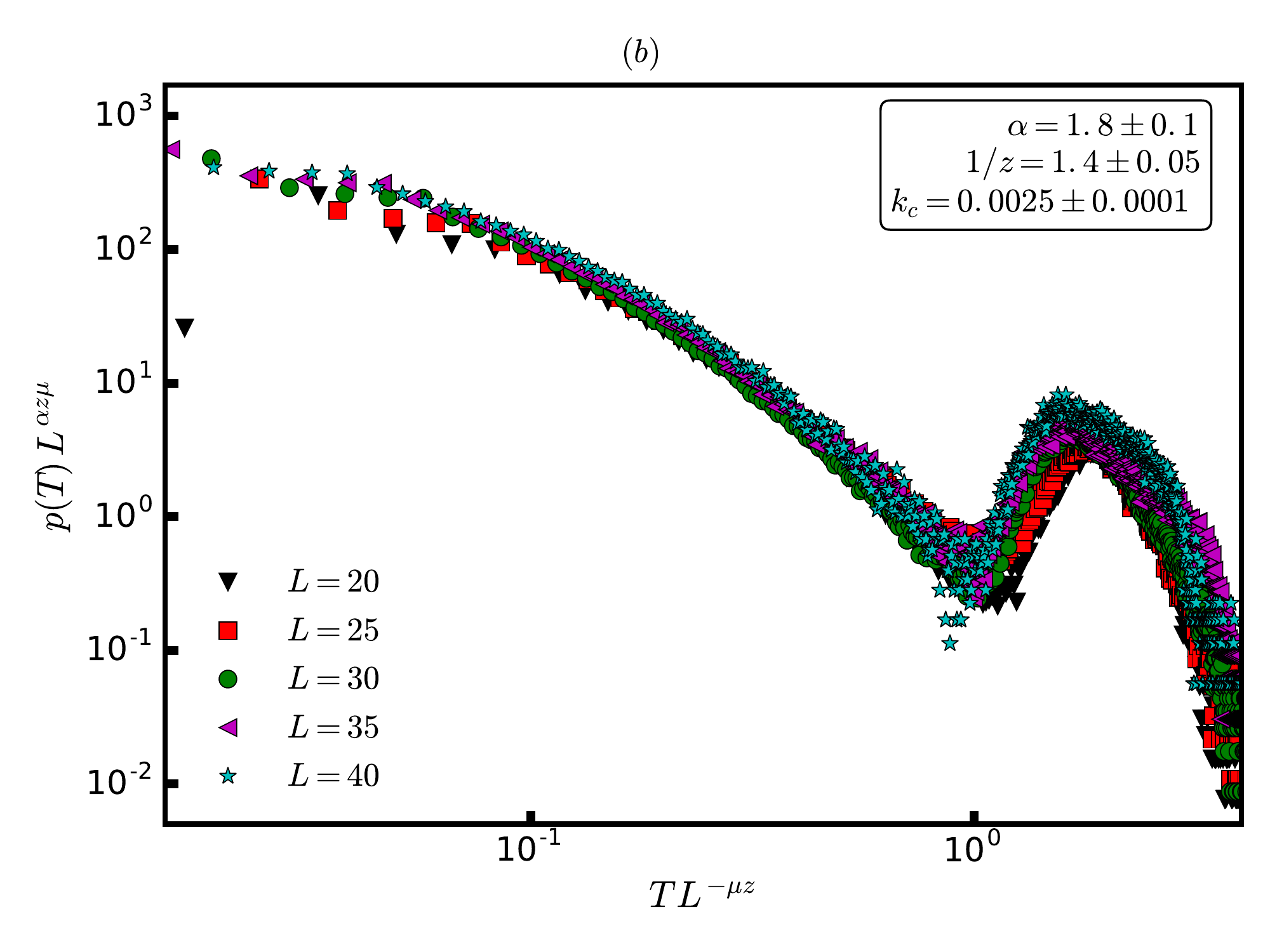}}
\resizebox{0.95\columnwidth}{!}{%
\includegraphics{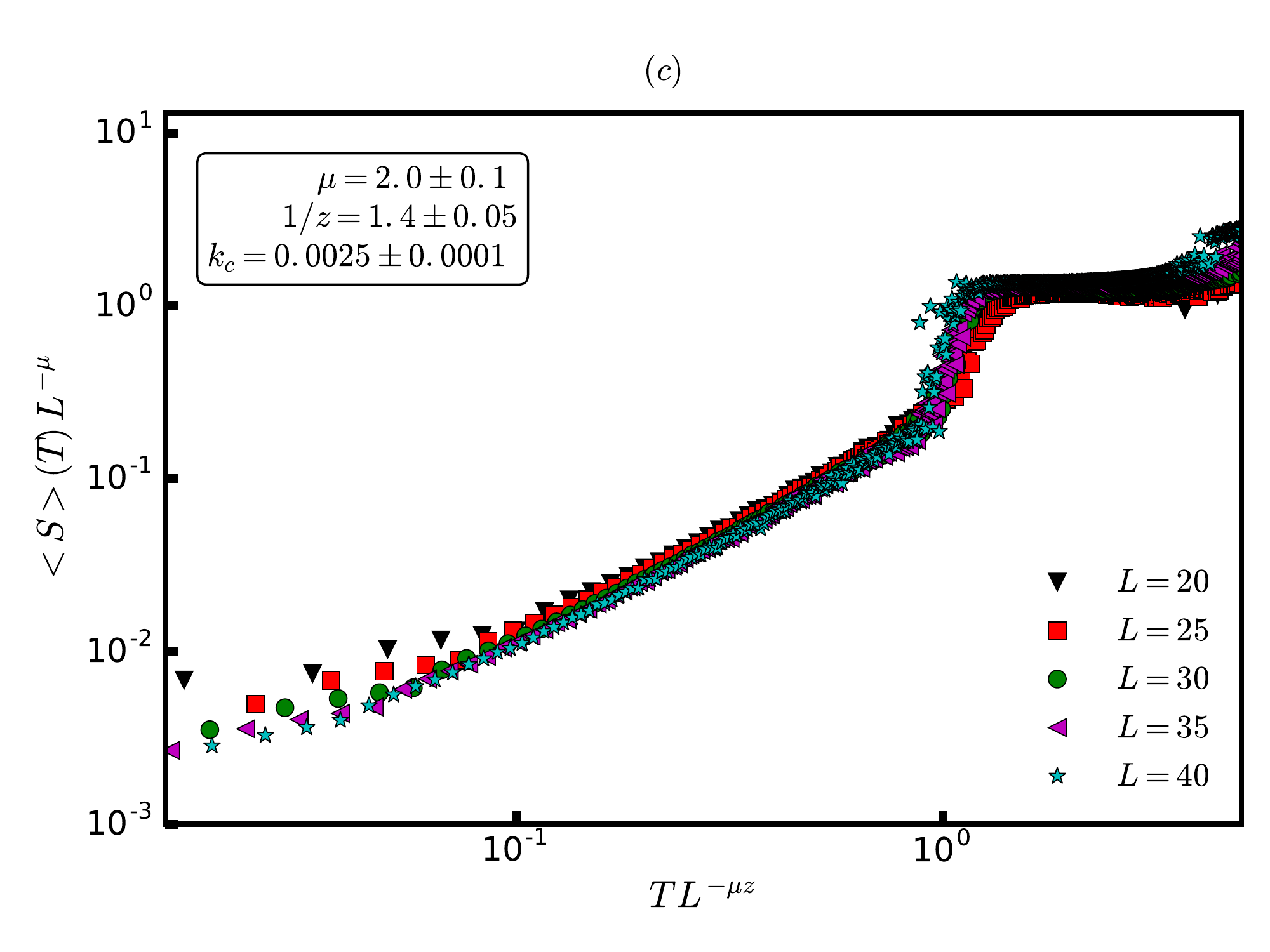}}
\caption{ (Color online) Collapse of the avalanche data shown in Fig~\ref{fig:avalanches}, using the scaling relations \eqref{fs}.}
\label{fig:avalanches_scale}
\end{figure}

\noindent The scaling theory requires the following relation between the exponents~ ~\cite{Friedman2012}
\begin{figure}
\centering
\resizebox{\columnwidth}{!}{%
\includegraphics{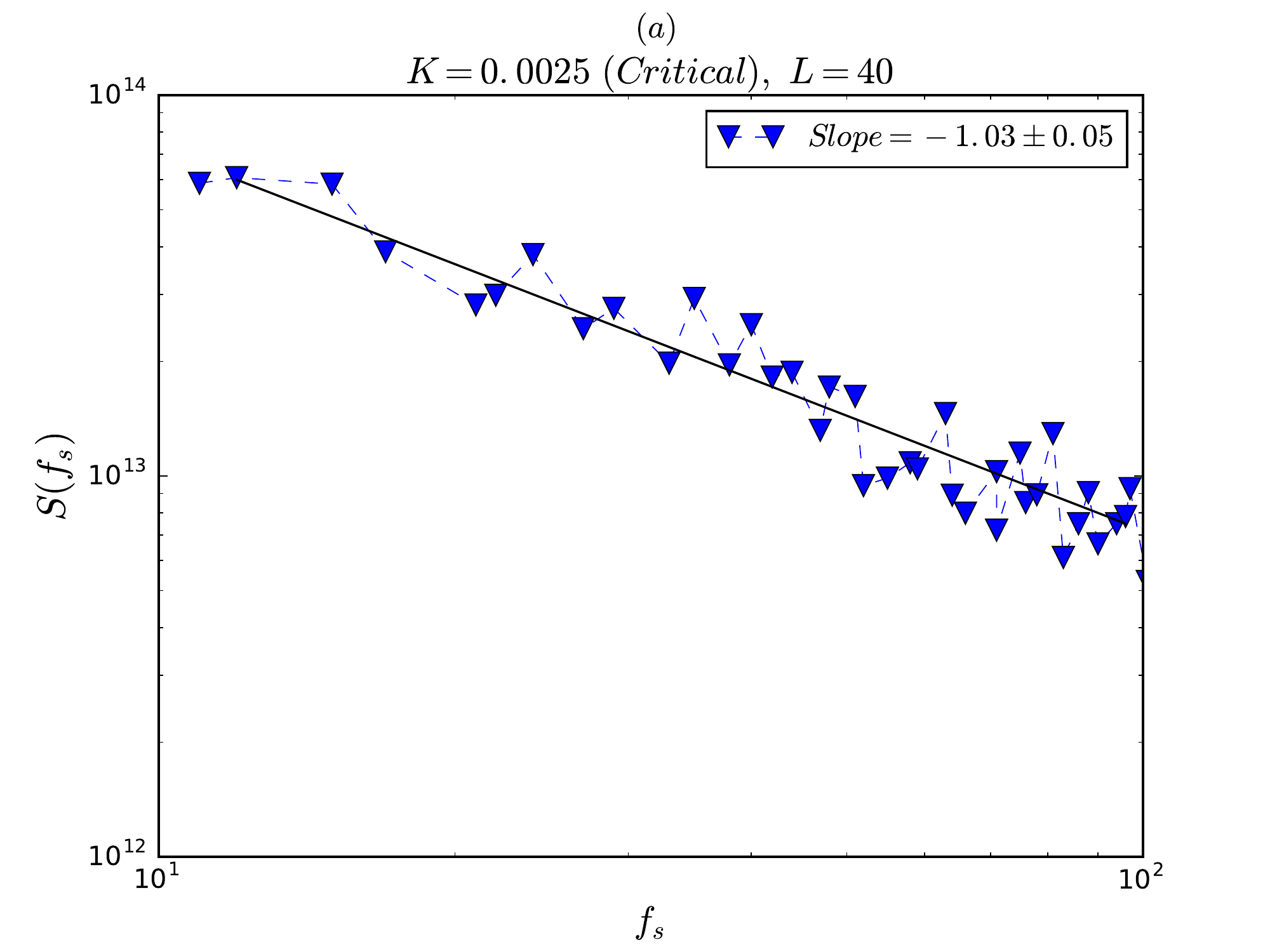}}
\resizebox{\columnwidth}{!}{%
\includegraphics{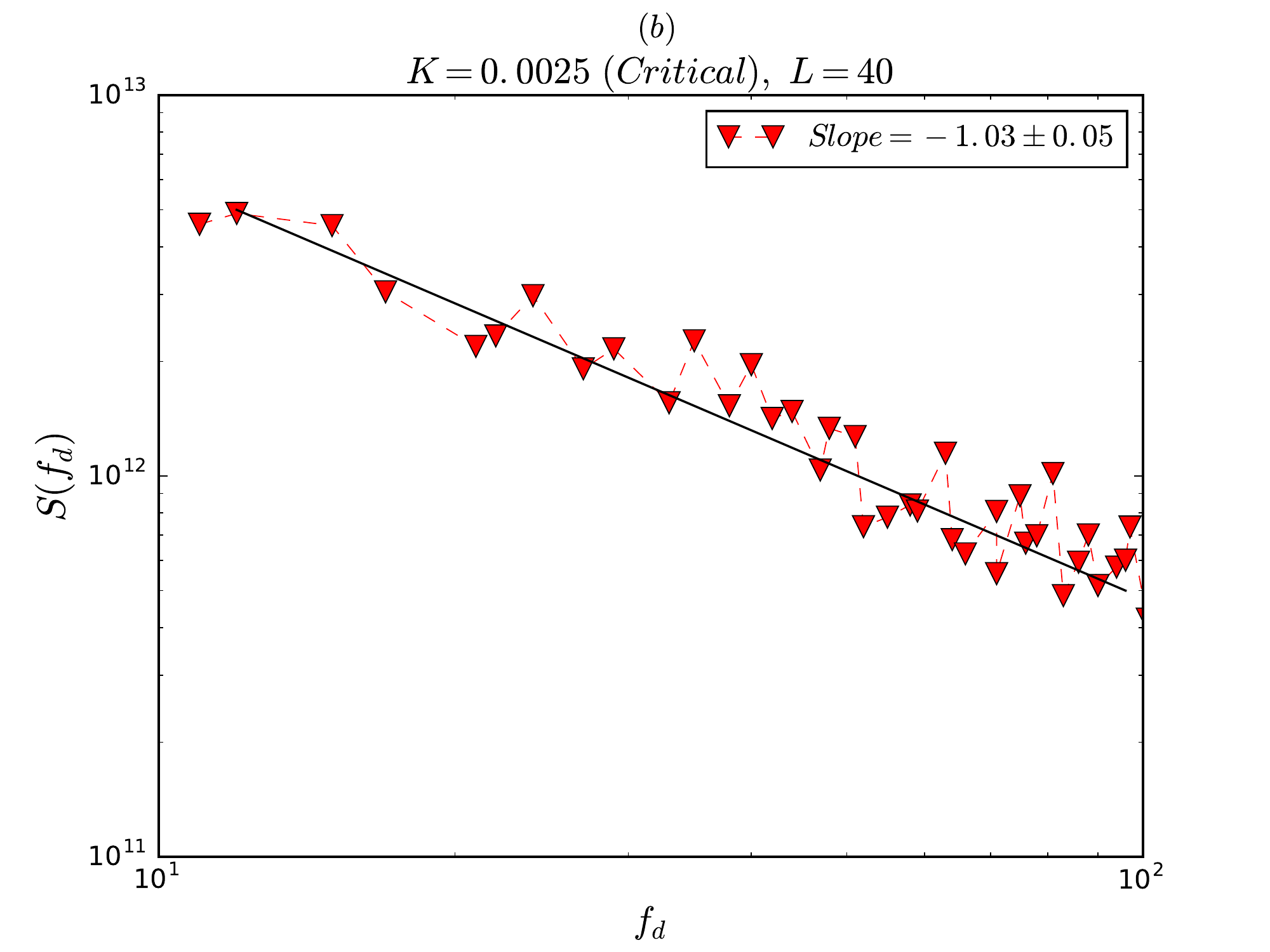}}
\caption{ (Color online) Logarithmic plot of the power spectra for (a) avalanche size and (b) avalanche duration time series at the critical cooperation $K_c=0.0025$.}
\label{fig:1/f}
\end{figure}

\begin{equation}
\label{s}
\frac{\alpha-1}{\tau-1} = \frac {1}{z}.
\end{equation}
\noindent The mean field prediction for the scaling exponents are $\tau=3/2, \alpha=2.0$, and $1/{z}= 2.0$~\cite{James}.

To determine the neuronal avalanches, we count the number of spikes in successive time steps until a silent period of equal or more than 5 time steps ($\delta t = 5$). Avalanche duration is considered to be the interval of the neuronal activity between any two silent periods and the number of spikes during this activity indicates the avalanche size (see Fig.~\ref{fig:avalanche_def}). 
In our simulation, we find that the neuronal avalanches show prominent scaling behavior at $K_c=0.0025$.
Fig.~\ref{fig:avalanches} illustrates the logarithmic relationship between the probability density functions of avalanche size and duration, and also the dependence of the average avalanche size upon the duration at $K=0.0025$ for the lattices with the linear size of $L=20,25,30,35,40$.
The data shown in this figure are rescaled based on the relations \eqref{fs} to properly collapse to a single plot. Fig.~\ref{fig:avalanches_scale} represents such a data collapse which is achieved by $\mu=2.0\pm 0.1$, $\tau=1.6\pm 0.1$, $\alpha=1.8 \pm 0.1$ and $1/z=1.4 \pm 0.05$ and $K_c=0.0025 \pm 0.0001$. It can be seen that the scaling relation \eqref{s} is satisfied with the simulation precision. The value of $\mu=2.0$ indicates that the size of the avalanches grows proportionally with the lattice size. Our estimation for the critical exponents $\tau$, $\alpha$ and $1/z$ are in fair agreement with the recent experimental values obtained in the primary visual cortex of both anesthetized and freely moving mice, while in critical states~\cite{fontenele2019criticality}. 

To close this section, we calculate the power spectra of avalanche size and avalanche duration time series generated according to the explanation above and shown in Fig.~\ref{fig:avalanche_def}. The results in logarithmic scale are sketched in Fig.~\ref{fig:1/f}, indicating the $1/f$ scaling of these time series. 

\section{Conclusion\label{conclusion}}

Neural avalanches are widely accepted as the hallmark of critical neural dynamics in the cortex, widely observed both {\it in vivo} and {\it in vitro}. However, it is not yet clear how local cortical networks converge on criticality. In our previous works on LIF, we showed a transitory neuronal behavior from an asynchronous to a fully synchronized regime, where a complex critical regime is detected between those two. We employed both temporal complexity and neural avalanche approaches and showed that they both meet at criticality. However, the existence of an order parameter to manifest this behavior and complement our model has been missing. Therefore, we set out to address this question by exploring the transitionary behavior of LIF neurons residing on a regular lattice of different sizes. We based our analysis on exploring the coupling of spikes to the phase of PAV.

We found that in a LIF model the coupling of spike timing and the phase of PAV, quantified by PLV, acts as an order parameter in our model. We showed that the increase in inter-neuronal cooperation parameter $K$, gives rise to a continuous phase transition from random phase irregular spiking (subcritical) to synchronous spiking (supercritical) with the spike timer, coupled tightly to the phase of PAV. This finding is in line with the hypothesis that the strength of SPC might be a control parameter that is used to harness the information processing capacity of the neural system frequently manifested in the form of neural avalanches in many empirical studies. Using the finite-size scaling analysis, we estimated the critical exponents corresponding to this transition and show that this transition does not belong to the universality class of either mean-field directed percolation nor directed percolation in $2+1$ dimensions. We also observed that at this critical point, the neuronal avalanches show power-law behavior with the scaling exponents close to the experimental values. This work suggests that the criticality observed in the neuronal systems could be the critical point of a synchronization transition.

Moreover, $1/f$ scaling in the power spectra of the avalanche time series suggests a potential connection between the neuronal avalanches and LFP signals. However, having known that the origin of LFP is the synaptic currents, our simple LIF model which lacks a dynamic for the synaptic current, does not allow us to extract LPFs. To this end, and to incorporate the neuronal synaptic dynamics, we need to modify our model by adding an extra-synaptic compartment to the LIF neurons. This will enable us to calculate {\it spike-LFP phase coupling}, which is the subject of our future research.

\section{acknowledgment}
\noindent {All authors gratefully acknowledge financial support from their affiliated institutions.}

\bibliographystyle{apsrev4-1}

\bibliography{biblio}

\end{document}